\documentclass[oneside,twocolumn,pra]{revtex4}

\usepackage{graphicx,amsfonts}

\graphicspath{/Users/Richard/Pictures/}
\usepackage[english]{babel}
\usepackage{amsmath}
\usepackage{graphicx}
\usepackage{bbold}
\usepackage{amssymb}
\usepackage[titletoc]{appendix}
\usepackage{verbatim}
\usepackage{setspace}
\newcommand{\etal}{{\it et al.\ }}

\newcommand{\ket}[1]{|#1\rangle}
\newcommand{\bra}[1]{\langle#1|}  

\begin{document}
\title{Entanglement Concentration with Quantum Non Demolition Hamiltonians}
\author{Richard Tatham}
\email{rt264@st-andrews.ac.uk}
\author{Natalia Korolkova}
\email{nvk@st-andrews.ac.uk}
\affiliation{School of Physics and Astronomy, University of St Andrews, Fife, Scotland}
\begin{abstract}
We devise and examine two procrustean entanglement concentration schemes using Quantum Non-Demolition (QND) interaction Hamiltonians in the continuous variable regime, applicable for light, for atomic ensembles or in a hybrid setting. We thus expand the standard entanglement distillation toolbox to the use of a much more general, versatile and experimentally feasible interaction class. The first protocol uses Gaussian ancillary modes and a non-Gaussian post-measurement, the second a non-Gaussian ancillary mode and a Gaussian post-measurement.  We explicitly calculate the density matrix elements of the non-Gaussian mixed states resulting from these protocols using an elegant Wigner-function based method in a numerically efficient manner. We then quantify the entanglement increase calculating the Logarithmic Negativity of the output state and discuss and compare the performance of the protocols.
\end{abstract}
\maketitle
\section{Introduction}

One of the biggest hurdles quantum communication has yet to overcome is the difficulty of generating highly entangled states between distant nodes. Due to the inherent noise in any optical setup, in reality it is only ever possible to generate partially entangled states at distant locations. The remedy to this problem is entanglement distillation \cite{Gisin1996,Bennett1996,Bennett1996a}. Local operations and classical communication (LOCC) are used to distil at least one maximally entangled bipartite state from multiple weakly entangled states. 

One of the latest developments in quantum information science is the use of infinite-dimensional systems or continuous variables (CV), such as the amplitude and phase quadratures of optical modes or the collective spin of macroscopic atomic ensembles. CV systems offer the advantage of cheap resources, easy generation and control, and off-the-shelf telecommunication components can be used for information processing. The downside of it is that it is in principle impossible to have maximally entangled states, and consequently there is an additional source of imperfections. Furthermore, finding concentration schemes in the continuous variable regime is made particularly difficult by a no-go theorem that states that distillation of Gaussian states cannot be carried out using Gaussian operations alone \cite{Eisert2002,Giedke2002,Fiurasek2002}. One has to resort to challenging non-Gaussian operations, like strong high-order nonlinear interactions or photon counting.

It is worth noting that in the continuous variable regime, more or less all of the proposed schemes for entanglement distillation contain variations of a common theme: beamsplitter interactions. On the theory side, Opartn\'y \etal \cite{Opartny2000} showed how beamsplitters and photon subtraction can be used to increase the entanglement in a two mode squeezed vacuum (TMSV) as part of a teleportation scheme. The distillation aspect of this was seized upon and improved \cite{Olivares2003,Eisert2004,Lund2009,Tatham2011} and a rigorous theoretical description was given by Kitagawa \etal \cite{Kitagawa2006}. Experimentally, there have been some notable successes \cite{Ourjoumtsev2007,Takahashi2010}, including with non-Gaussian noise \cite{Dong2010}. In a fairly recent proposal \cite{Walmsley2011} a complete CV entanglement ``distillery'' has been proposed, harnessing limited physical space for storing quantum states and distilling entanglement. There, clever manipulation of an imperfect quantum memory complements the beamsplitter based entanglement concentration scheme.

However, the net could be cast wider. Theoretical work has been carried out utilizing other interactions between an entangled state and ancillary modes, notably the cross-Kerr effect \cite{Fiurasek2003a,Menzies2006}, in which the nonlinear interaction plays the role of the necessary non-Gaussian element, and weak values \cite{Menzies2007}. Still, very little is known about entanglement distillation protocols using so-called Quantum Non-Demolition (QND) interaction Hamiltonians, although they lend themselves readily to many experimental systems where distillation could be useful. For example the off-resonant dipole interaction between heavily polarized light and the macroscopic spin properties of an atomic ensemble can be modeled as QND and two atomic ensembles were successfully entangled, albeit weakly, in this way \cite{Julsgaard2001}. Furthermore, for light it was also shown that it is possible to create a QND Hamiltonian using optical instruments such as biased beamsplitters and squeezers in combination \cite{Braunstein2005a}.

The Quantum Non-Demolition interactions \cite{QNDreview,GrangierReview} are well known in quantum optics for quantum metrology, entanglement generation and other purposes. They are described by Hamiltonians of the form e.g.
\begin{equation}
\hat{H}_{\textrm{int}}=\kappa\hat{P}_S\hat{P}_A
\label{QND}
\end{equation}
where $\kappa$ is the interaction strength and $\hat{P}_{S/A}$ are the momentum quadratures of the system and ancillary mode. The system and ancillary mode interact in such a way as to leave one quadrature component of each subsystem intact whilst phase shifting the conjugate components.

In this paper, we suggest and explore two procrustean entanglement distillation schemes that use QND interactions, and quantify their performance showing the (probabilistic) increase of the the logarithmic negativity of a two mode squeezed vacuum, defined in the Fock basis as
\begin{equation}
\left|\Psi\right\rangle=\sqrt{1-\lambda^2}\sum^{\infty}_{n=0}\lambda^n\left|nn\right\rangle
\label{TMSSFock}
\end{equation}
where $\ket{n}$ is the photon number and $\lambda=\tanh r$ is the squeezing. For the quantitative analysis of these schemes, we amalgamate ideas on using the well-known formalism of Gaussian states to represent non-Gaussian states \cite{Garcia2005} (used more recently in \cite{Zhang2011}) and calculate density matrix elements using multivariate Hermite polynomials \cite{Dodonov1994,Adam1995,Manko1998}. Our approach is incredibly versatile, making additional interactions and double passes of the interacting modes as simple as multiplying $8\times8$ matrices and allowing for detector inefficiencies to be considered simply. Hence numerical calculations are made fast and accurate. Thus the novelty of the proposed approach to distillation lies both in the scheme design (QND-based) and in the methodology of the calculations. Notably, the suggested scheme is not bound to a certain physical system  and can be applied to the distillation of entanglement in light modes, atomic ensembles or in hybrid systems, the essential requirement being solely the form of the Hamiltonian underlying the interaction.

The paper is organized as follows. In section \ref{prelim} we remind the reader of the important properties of Gaussian states required. In section \ref{protone} we study the first scheme, a procrustean entanglement concentration protocol replacing the conventional beamsplitter interaction with the QND interaction between a Gaussian ancilla and a Gaussian entangled state. More precisely, we calculate the logarithmic negativity of the non-Gaussian mixed state resulting from the QND interaction between the two mode squeezed vacuum and a Gaussian ancillary mode subsequently subjected to a non-Gaussian measurement. In section \ref{prottwo}, we explore an alternative scheme where the ancillary mode is non-Gaussian and the post-measurement is Gaussian. In both cases the entanglement of the system is shown to increase demonstrating a successful continuous variable entanglement distillation based on the QND Hamiltonian.

\section{Preliminaries}
\label{prelim}
An N-mode Gaussian state can be completely defined by the first and second order moments of the quadrature operators $\hat{R}_j$ with $\textbf{R}=\left(x_1,p_1,\ldots,x_N,p_N\right)^T$. The Wigner function for such a state is given by
\begin{equation}
\mathcal{W}\left(x_1,p_1,\ldots,x_N,p_N\right)=\frac{\exp\left[-\left(\textbf{R}^T-\textbf{d}^T\right)\gamma^{-1}\left(\textbf{R}-\textbf{d}\right)\right]}{\pi^N\sqrt{\det\gamma}}
\label{WignerGauss}
\end{equation}
where $\textbf{d}$ is the displacement vector $\textbf{d}=\left(\left\langle x_1\right\rangle,\left\langle p_1 \right\rangle,\ldots,\left\langle x_N \right\rangle,\left\langle p_N \right\rangle\right)^T$ and $\gamma$ is the covariance matrix of the state in question, defined as
\begin{equation}
\gamma_{lm}=\left\langle\hat{R}_l\hat{R}_m+\hat{R}_m\hat{R}_l\right\rangle-2d_ld_m.
\end{equation}
The two mode squeezed state that we wish to distil has covariance matrix
\begin{equation}
\gamma^{TMSS}_{AB}=\left(\begin{array}{cccc} \cosh(2r)&0&\sinh(2r)&0\\0&\cosh(2r)&0&-\sinh(2r)\\\sinh(2r)&0&\cosh(2r)&0\\0&-\sinh(2r)&0&\cosh(2r)\end{array}\right).
\end{equation}
This Gaussian distribution is centred at the origin in phase space and so $\textbf{d}$ of equation \eqref{WignerGauss} is the zero vector. The squeezing parameter is denoted $r$.

As a well-known consequence of the Stone-Von-Neumann theorem, any quadratic unitary operation at the Hilbert space level corresponds to a symplectic operation at the phase space level, that preserves the symplectic form $\Omega=\oplus^N_{j=1}i\sigma_y$. In what follows, the following symplectic operations are relevant. Firstly, the QND interaction between two modes (system and ancilla) can be described by its action on phase space, e.g.
\begin{equation}
S^{\left(\kappa\hat{P}\hat{P}\right)}_{QND}=\left(\begin{array}{cccc}1&0&0&\kappa\\0&1&0&0\\0&\kappa &1&0\\0&0&0&1\\\end{array}\right)
\end{equation}
corresponds to the Hamiltonian of the form given in Eq.~(\ref{QND}). Similarly,
$S^{\left(\kappa\hat{X}\hat{X}\right)}_{QND}$, $S^{\left(\kappa\hat{X}\hat{P}\right)}_{QND}$, and $S^{\left(\kappa\hat{P}\hat{X}\right)}_{QND}$ can also be defined where  $\hat X$ is the position quadrature. The phase shift of one mode is defined as
\begin{equation}
S_{PH}\left(\theta\right)=\left(\begin{array}{cc}\cos\theta & \sin\theta\\-\sin \theta & \cos\theta\\\end{array}\right)
\label{Sphase}
\end{equation}
where $\theta$ is the angle through which the Gaussian mode is rotated in phase space. We also define the single mode squeezer
\begin{equation}
S_{SQ}\left(s\right)=\left(\begin{array}{cc}e^s&0\\0&e^{-s}\\\end{array}\right)
\end{equation}
which yields $\gamma_{SQ}=\textrm{diag}\left(e^{2s},e^{-2s}\right)$ when applied to the vacuum covariance matrix $\gamma_{vac}=\mathbb{1}$ . If $s>0$ then the mode is squeezed in $p$ and the mode is squeezed in $x$ if $s<0$.

Noisy quantum channels and phase insensitive amplifiers can not be modeled by Gaussian unitary transformations. One of the biggest losses by detectors is when they erroneously fail to register a detection. For this reason we model inefficient detectors as being perfect detectors behind a beamsplitter of transmittance $\eta$ (see e.g. \cite{Nemoto2002,Eisert2004}). This can be seen as combining the mode with the vacuum on a beamsplitter and tracing out the vacuum. Such an operation transforms the covariance matrix as
\begin{equation}
\gamma\rightarrow S_{\textrm{loss}}\left(\eta\right)\gamma S^T_{\textrm{loss}}\left(\eta\right)+\left(1-\eta\right)\mathbb{1}
\end{equation}
where $S_{\textrm{loss}}\left(\eta\right)=\sqrt{\eta}\mathbb{1}$.

\section{Protocol I: Gaussian ancillary mode and non-Gaussian measurement}
\label{protone}

\begin{figure}
\includegraphics[width=7cm]{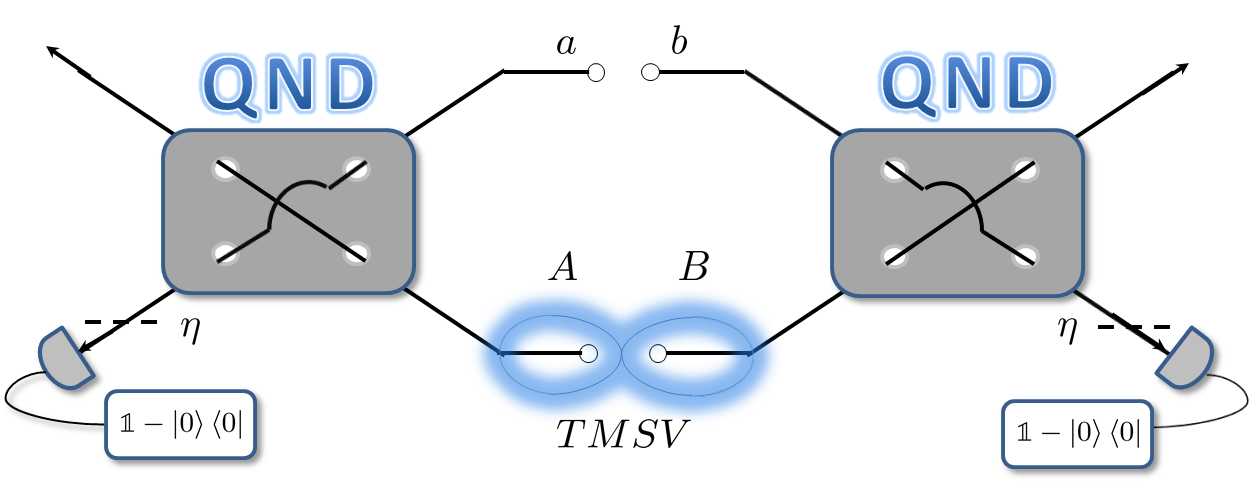}
\caption{\small The two modes of a two mode squeezed vacuum interact with two vacuum modes which are subsequently measured for the presence or absence of photons. If successful, the entanglement of the two mode squeezed state increases.}
\label{prot1}
\end{figure}
The first protocol, depicted in Figure \ref{prot1}, facilitates the concentration of the entanglement of some continuous degree of freedom coupled to ancillary modes with QND interactions. 
%The first protocol depicted in Figure \ref{prot1} is the amelioration of our previous protocol, where the distillation of entanglement between two atomic ensembles is performed approximating the QND interaction by an effective light-atom beamsplitter \cite{Tatham2011}.
%The current, generalized protocol facilitates the concentration of the entanglement of some continuous degree of freedom coupled to ancillary light modes with  QND interactions.
The entangled state is in the form (\ref{TMSSFock}), where the exact meaning of the basis states $\vert n n\rangle$ is determined by the used system. The ancillary modes are in a vacuum state and so are described in phase space by a Gaussian Wigner function. The required non-Gaussian element  \cite{Eisert2002,Giedke2002,Fiurasek2002} is in this case fulfilled by a non-Gaussian measurement. %We require then a non-Gaussian element for entanglement distillation  \cite{Eisert2002,Giedke2002,Fiurasek2002} and in this scheme it is represented by a non-Gaussian measurement.
Thus we begin with a two mode squeezed vacuum (modes $A$ and $B$) and two vacuum ancillary modes ($a$ and $b$) (Figure \ref{prot1}). The initial covariance matrix is given by $\gamma_{\textrm{init}}=\gamma^{TMSS}_{AB}\oplus\gamma_{vac,a}\oplus\gamma_{vac,b}$. After the desired QND interaction is performed between $A$ and $a$, and $B$ and $b$, the two ancillary modes are detected for the presence or absence of photons on a detector of efficiency $\eta$. That is, the non-Gaussian measurement $\hat{\Pi}=\mathbb{1}-\ket{0}\bra{0}$ is performed on modes $a$ and $b$. Immediately prior to the measurement, the covariance matrix of all four modes can be described by
\begin{eqnarray}
&\gamma=\nonumber\\
&\left(\mathbb{1}_{AB}\oplus S_{\textrm{loss},ab}\left(\eta\right)\right)S_{\textrm{QND}}\,\gamma_{\textrm{init}}\,S^T_{\textrm{QND}}\left(\mathbb{1}_{AB}\oplus S_{\textrm{loss},ab}\left(\eta\right)\right)\nonumber\\
%\quad
&+\left(\mathbb{0}\oplus\left(1-\eta\right)\mathbb{1}_{ab}\right).
\label{gammaqnd}
\end{eqnarray}
Conceivably, other interactions could be considered by modifying \eqref{gammaqnd}, such as double-pass schemes (see \cite{Tatham2011,polzik-multipass}). After the measurement, the total Wigner function of modes $A$ and $B$ is given by
\begin{equation}
\mathcal{W}_{AB}=M\sum^1_{i,j=0}\left(-1\right)^{i+j}P^{(ij)}\mathcal{W}^{(ij)}_{AB}.
\label{Wignertot}
\end{equation}
That is, the total Wigner function $\mathcal{W}_{AB}\left(x_A,p_A,x_B,p_B\right)$ is given by a superposition of normalized Gaussian Wigner functions $\mathcal{W}^{(ij)}_{AB}\equiv\mathcal{W}^{(ij)}_{AB}\left(x_A,p_A,x_B,p_B\right)$ with $i=0$ meaning that the mode $a$ has been projected onto $\left|0\right\rangle\left\langle0\right|$ and $i=1$ implies that $a$ has been projected onto $\mathbb{1}$ (i.e. traced out). The index $j$ tells what has happened to mode $b$. The constants $P^{(ij)}$ are the probabilities associated with $\mathcal{W}^{(ij)}_{AB}$ and $M$ is a global normalisation constant given as $M=\left(\sum_{i,j}{\left(-1\right)^{i+j}P^{(ij)}}\right)^{-1}$.

The $4\times4$ covariance matrices $\gamma^{(ij)}_{AB}$ that make up the Gaussian Wigner functions $\mathcal{W}^{(ij)}_{AB}$ and their normalisation constants $P^{(ij)}$ are calculated from $\gamma$ by considering the Wigner Overlap formula. Tracing out a mode corresponds to an overlap with $\mathcal{W}_{\mathbb{1}}=\left(1/2\pi\right)$ and a projection onto the vacuum corresponds to an overlap with $\mathcal{W}_0=\pi^{-1}\exp[-x^2-p^2]$. For further details see e.g. \cite{Garcia2005}.

We define
\begin{equation}
\Gamma=\gamma^{-1}=\left(\begin{array}{cc}\Gamma_{AB}&\sigma\\\sigma^T&\Gamma_{ab}\\\end{array}\right)
\end{equation}
where $\Gamma_{AB}$ describes the entangled modes $A$ and $B$ and $\Gamma_{ab}$ describes the ancilla modes $a$ and $b$ with $\sigma$ capturing the cross correlations between entangled and ancilla modes. Then $\gamma^{(ij)}_{AB}=\left(\Gamma^{(ij)}_{AB}\right)^{-1}$ where
\begin{equation}
\Gamma^{\left(ij\right)}_{AB}=\Gamma_{AB}-\sigma\left(\Gamma^{\left(ij\right)}_{ab}\right)^{-1}\sigma^T, \nonumber
\end{equation}
\begin{equation}
\Gamma^{\left(ij\right)}_{ab}=\Gamma_{ab}+\left(1-i\right)\mathbb{1}\oplus\left(1-j\right)\mathbb{1}.
\end{equation}
The normalisation constants $P^{\left(ij\right)}$ can be calculated by the rather straightforward integration of $\mathcal{W}^{\left(ij\right)}_{AB}$ which can simply be thought of as the overlap of modes $A$ and $B$ with $\mathcal{W}_{\mathbb{1}}$. Consequently,
\begin{equation}
P^{(ij)}=\frac{2^{2-i-j}\sqrt{\det\Gamma}}{\sqrt{\det\Gamma^{\left(ij\right)}_{AB}}\sqrt{\det\Gamma^{\left(ij\right)}_{ab}}}.
\end{equation}
At this point, we have specified all the quantities in the expression \eqref{Wignertot} for the Wigner function of 
the two mode squeezed state after QND interactions with vacuum modes that are later detected for the presence or absence of photons on inefficient detectors. Hence we are able to describe the entangled state modified by the QND interaction by use of the Wigner function \eqref{Wignertot}. However, to show that the entanglement in the two modes is increased, we require the density matrix elements of the state. These can be calculated with the help of the Q function and multidimensional Hermite polynomials \cite{Dodonov1994,Adam1995,Manko1998}.

The Q function is defined as the convolution of the Wigner function with the vacuum \cite{Essential} and so
\begin{equation}
\mathcal{Q}^{(ij)}_{AB}\left(\alpha,\beta\right)=\frac{\exp\left[-\mathcal{R}^\dagger\left(\textbf{U}\gamma^{(ij)}\textbf{U}^\dagger+\mathbb{1}\right)^{-1}\mathcal{R}\right]}{\pi^2\sqrt{\det\left(\gamma^{(ij)}+\mathbb{1}\right)}}
\end{equation}
where $\mathcal{R}=\left(\alpha,\alpha^\dagger,\beta,\beta^\dagger\right)^T$ and $\textbf{U}$ is the unitary transformation between $\left(\hat{x}_A,\hat{p}_A,\hat{x}_B,\hat{p}_B\right)^T\rightarrow\left(\hat{a},\hat{a}^\dagger,\hat{b},\hat{b}^\dagger\right)^T$:
\begin{equation}
\textbf{U}=\frac{1}{\sqrt{2}}\left(\begin{array}{cccc}1&i&0&0\\1&-i&0&0\\0&0&1&i\\0&0&1&-i\end{array}\right).
\end{equation}
The Q function then acts as a generating function for the density matrix elements, $\rho^{(ij)}_{lmnp}$.
\begin{eqnarray}
\rho^{(ij)}_{lmnp}=&\left\langle l\right|\left\langle m\right|\hat{\rho}^{(ij)}\left| n \right\rangle\left| p \right\rangle\nonumber\\ =&\frac{\left(2\pi\right)^2}{\sqrt{l!m!n!p!}}\left(\frac{\partial^{l+n}}{\partial\alpha^{*l}\partial\alpha^n}\right)\left(\frac{\partial^{m+p}}{\partial\beta^{*m}\partial\beta^p}\right)\nonumber\\&\quad\times{\left[\mathcal{Q}^{(ij)}_{AB}\left(\alpha,\beta\right)e^{\left|\alpha\right|^2+\left|\beta\right|^2}\right]\vline}_{\alpha=\beta=0}
\end{eqnarray}
The matrix element is most easily expressed in terms of the four dimensional Hermite polynomials $H_{lmnp}$ (see appendix \ref{Hemitepolys}). After successful detection the total density matrix is given by $\rho$ with matrix elements
\begin{equation}
\rho_{lmnp}=\frac{4M}{\sqrt{l!m!n!p!}}\sum^1_{i,j=0}\frac{\left(-1\right)^{i+j}P^{(ij)}H^{\{\textbf{C}^{(ij)},0\}}_{lmnp}}{\sqrt{\det\left(\gamma^{(ij)}+\mathbb{1}\right)}}
\label{denstot}
\end{equation}
where for brevity we have defined $H^{\{\textbf{C}^{(ij)},0\}}_{lmnp}\equiv H^{\{\textbf{C}^{(ij)},0\}}_{lmnp}\left(0,0,0,0\right)$. The matrices $\textbf{C}^{(ij)}$ are derived from the matrix equation
\begin{equation}
\textbf{C}^{(ij)}=\textbf{B}\left[\left(\textbf{U}\gamma^{(ij)}\textbf{U}^\dagger+\mathbb{1}\right)^{-1}-\frac{1}{2}\mathbb{1}\right]\textbf{D}
\end{equation}
with 
\begin{equation}
\textbf{B}=\left(\begin{array}{cccc}1&0&0&0\\0&0&1&0\\0&1&0&0\\0&0&0&1\\\end{array}\right),\quad\textbf{D}=\left(\begin{array}{cccc}0&0&1&0\\1&0&0&0\\0&0&0&1\\0&1&0&0\\\end{array}\right).
\label{swapit}
\end{equation}
The purpose of $\textbf{B}$ and $\textbf{D}$ is simply to rearrange the elements of the matrix in such a way as to make the matrix compatible with the form given in \eqref{Hermite} and to make $\rho_{lmnp}$ proportional to $H^{\{\textbf{C}^{(ij)}\}}_{lmnp}$. For single QND passes the matrix $\textbf{C}^{(ij)}$ is always real.

With the density matrix elements of modes $A$ and $B$ we can now show that the entanglement of the system has increased. To quantify this we use the additive Logarithmic Negativity \cite{Vidal2002} defined as
\begin{equation}
\mathcal{E}\left(\hat{\rho}_{AB}\right)=\ln\left\|\hat{\rho}^{PT}_{AB}\right\|=\ln\left|2\mathcal{N}\left(\hat{\rho}_{AB}\right)+1\right|
\label{logneg}
\end{equation}
where $\left\|\hat{\rho}^{PT}_{AB}\right\|$ is the trace-norm of the partial transpose of $\hat{\rho}_{AB}$. The Negativity, $\mathcal{N}$ is defined as the sum of the negative eigenvalues of $\left\|\hat{\rho}^{PT}_{AB}\right\|$.

\begin{figure}
\includegraphics[width=7cm]{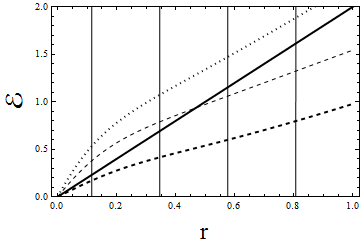}
\caption{Logarithmic Negativity of TMSV (solid line), and after a QND interaction of the form $\kappa\hat{X}\hat{P}$ with strength 0.1 (dotted), 0.5 (dashed), and 1 (thick dashed). The scheme used is that depicted in Figure \ref{prot1}. The entanglement increase is best when weak interactions are used. Strong interactions can decrease the entanglement in the initial TMSV. The vertical lines from left to right indicate where the initial squeezing $r$ is 1dB, 3dB, 5dB, and 7dB.}
\label{QND1}
\end{figure}

The logarithmic negativity of the mixed state resulting from this protocol is shown in Figure \ref{QND1}. 
The graph demonstrates the performance of the protocol in different regimes: depending on the initial entanglement content in the
two-mode squeezed vacuum and the QND interaction strength (see caption). Successful entanglement concentration is achieved for weak to medium values of $\kappa$, i.e. for a moderate QND coupling between the TMSV and ancilla modes, for a wide range
of the initial squeezing in the TMSV. If the interaction is too strong (e.g. $\kappa=1$) then the effect of the QND interaction on the entanglement content is negative, but as the interaction strength decreases, the logarithmic negativity of the resultant state increases and the performed QND operation is beneficial at least when the initial squeezing of the TMSV is weak. That is, as the interaction strength decreases, it becomes more likely that just a single photon is subtracted from each mode of the two mode squeezed vacuum and these photons are then detected. As expected, this compares well to the case where beamsplitters are used in place of QND interactions, investigated numerically in \cite{Kitagawa2006} for light TMSV and in \cite{Tatham2011} for TMSV-entangled atomic ensembles. After all, if the pure state resulting from a single photon subtraction in both modes of the entangled state is more entangled than the initial state, then it should not matter which interaction exactly is used to subtract the photons.

\section{Protocol II: Non-Gaussian ancillary mode and Gaussian measurement}
\label{prottwo}

\begin{figure*}[bth]
\begin{center}
\includegraphics[width=13cm]{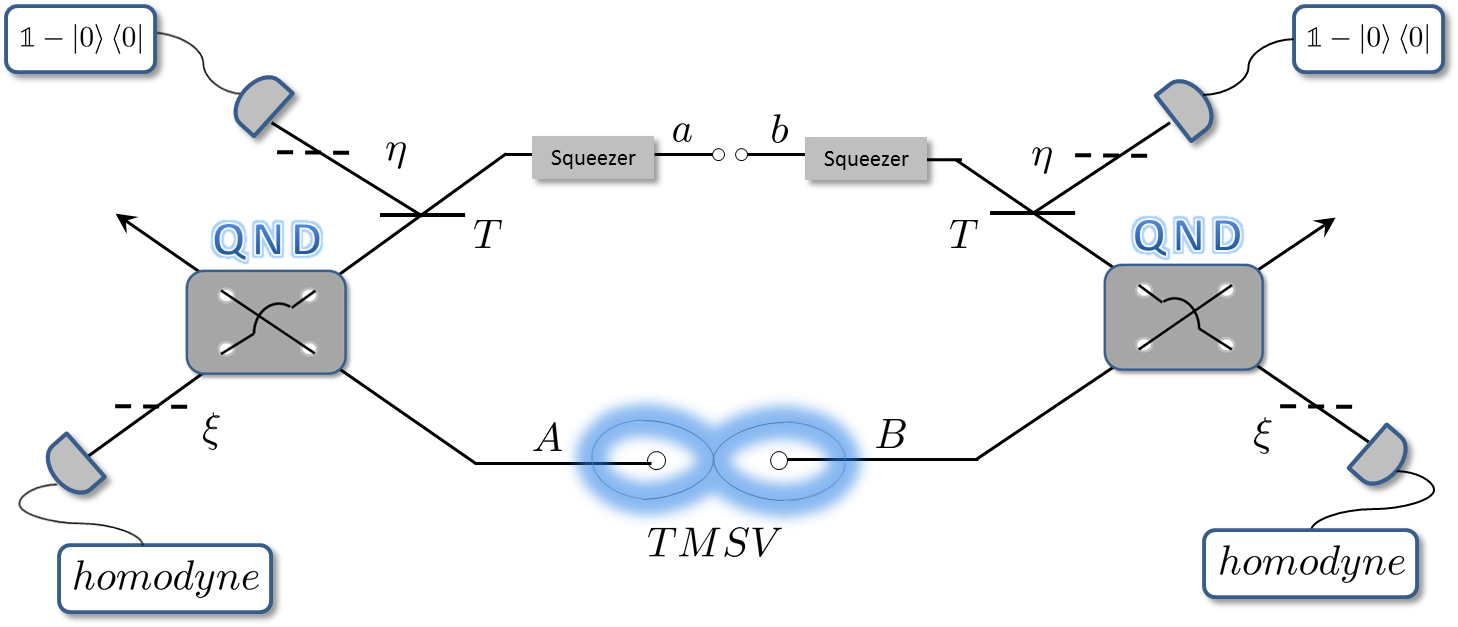}
\end{center}
\caption{\small A non-Gaussian ancillary state is first created in a preparatory step by subtracting photons from a squeezed vacuum. The ancilla then interacts with the TMSV via QND interactions and is measured by homodyne detection.}
\label{prot2}
\end{figure*}

Whereas Protocol I relies on a Gaussian ancillary mode interacting with the Gaussian TMSV via QND interaction, and a subsequent non-Gaussian measurement, we here consider a protocol which uses a non-Gaussian ancilla and a Gaussian measurement after the QND interaction (Figure \ref{prot2}). The advantage of this new protocol is that probabilistic non-Gaussian measurements are performed in the preparatory stage and, if successful, the QND interaction can be carried out in the next step. Also, the detection part of the scheme outperforms the first protocol as the Gaussian measurement, homodyne detection, can be highly efficient. The same methods as above can be used to calculate the logarithmic negativity of this scheme. 

\subsection{Preparatory step}

In the protocol of Figure~\ref{prot2} the ancillary modes $a$ and $b$ are in a quantum state with a non-Gaussian Wigner function. Specifically, they are in a photon-subtracted squeezed vacuum state. We consider first the preparation of the ancillary modes.  In contrast to the previous scheme the modes $a$ and $b$ are initially in a squeezed vacuum state, $\gamma_{ab}=\gamma_{SQ,a}\left(s\right)\oplus\gamma_{SQ,b}\left(s\right)$. The removal of photons from the mode is modeled in the following way. The modes $a$ and $b$ pass through a beamsplitter of transmittance $T^2$ and any subtracted photons are subsequently detected with efficiency $\eta$. As the transmittance $T^2$ increases, the model approaches the limit of just a single photon being subtracted from the mode. In what follows, in all numerical calculations we take $T=0.95$.

If photon subtraction is successful, the resultant non-Gaussian Wigner function becomes a superposition of Gaussian Wigner functions in the form $\mathcal{W}_{ab}=M_{ab}\sum_{i,j}\left(-1\right)^{i+j}P^{(ij)}\mathcal{W}^{(ij)}_{ab}$ with probability $M_{ab}=\left(\sum\left(-1\right)^{i+j}P^{(ij)}\right)^{-1}$. The corresponding covariance matrices that make up the Wigner functions $\mathcal{W}^{(ij)}_{ab}$ are given by $\gamma^{(ij)}_{ab}=\gamma^{(i)}_{\textrm{prep}}\oplus\gamma^{(j)}_{\textrm{prep}}$ where
\begin{equation}
\gamma^{(k)}_{\textrm{prep}}=\left(\begin{array}{cc}\vartheta_+(k)&0\\0&\vartheta_-(k)\\\end{array}\right)
\end{equation}
and
\begin{equation}
\vartheta_\pm(k)=1+\frac{\left(2-k\right)\left(e^{\pm2s}-1\right)T^2}{1+\left(1-k\right)\left(1+\eta\left(e^{\pm2s}-1\right)\left(1-T^2\right)\right)}.
\end{equation}
The coefficients $P^{(ij)}$ are calculated as
\begin{equation}
P^{(ij)}=\frac{2^{2-i-j}\sqrt{\vartheta_+(i)\vartheta_-(i)\vartheta_+(j)\vartheta_-(j)}}{\sqrt{\tau_+(i)\tau_-(i)\tau_+(j)\tau_-(j)}}
\end{equation}
with
\begin{equation}
\tau_\pm(k)=1-k-\frac{1+\left(e^{\pm2s}-1\right)T^2}{\left(1-\eta\right)\left(e^{\pm2s}-1\right)\left(1-T^2\right)-e^{\pm2s}}
\end{equation}
At this point we have a non-Gaussian state $\mathcal{W}_{ab}$, consisting of a superposition of Gaussian states. If the preparatory step could be performed independently and only passed to the interaction and homodyne measurement if successful then we could consider $\eta=1$.

\subsection{Interaction and Detection}
A non-Gaussian state produced in the preparatory step further interacts with the TMSV via QND coupling and the ancilla is subsequently passed on for homodyning on a detector of efficiency $\xi$. An angle $\theta$ describes the phase of the homodyne measurement, that is the angle at which homodyning is performed in phase space. An angle of $\theta=0$ corresponds to a measurement of position quadrature $x_{\theta=0}=x$ (the $x$ marginal distribution) whereas an angle of $\pi/2$ describes the momentum quadrature measurement $x_{\theta=\pi/2}=p$ (the p-distribution). Thus the homodyne measurement is characterised by the generalized quadratures $x_{\theta,a},x_{\theta,b}$. Directly before measurement, we can calculate the covariance matrices to be
\begin{eqnarray}
\gamma^{(ij)}_{ABab}=&\left(\mathbb{1}_{AB}\oplus S_{PH}\left(\theta\right)\right)\gamma^{(ij)}_{int}\left(\mathbb{1}\oplus\xi\mathbb{1}\right)\left(\mathbb{1}_{AB}\oplus S^T_{PH}\left(\theta\right)\right)\nonumber\\&+\left(1-\xi\right)S_{PH}\left(\theta\right)S^T_{PH}\left(\theta\right)
\end{eqnarray}
with
\begin{equation}
\gamma^{(ij)}_{int}=S_{\textrm{QND}}\left[\gamma^{TMSS}_{AB}\oplus\gamma^{(ij)}_{ab}\right]S^T_{\textrm{QND}}.
\end{equation}
To see what happens after the homodyne detection we project the quadratures $x_{\theta,a}$ and $x_{\theta,b}$ onto an outcome $z$. We also trace out the conjugate quadratures ($x_{\theta+\pi/2,a}$ and $x_{\theta+\pi/2,b}$).

\begin{figure}[tb]
\begin{center}
\includegraphics[width=7cm]{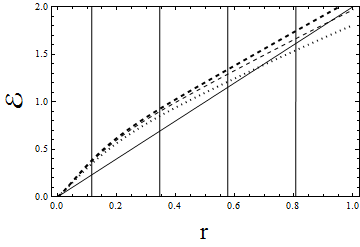}
\end{center}
\caption{Performance of protocol II (Fig.~\ref{prot2}): Logarithmic Negativity of TMSV before (solid line), and after a QND interaction for different initial squeezing of the ancilla modes. The QND interaction is of the form $\kappa\hat{X}\hat{P}$ and has strength $\kappa=0.5$; homodyne outcome $z\approx0$. The ancillary modes are squeezed by $s=0.1 (0.87dB)$ (black dotted), $s=0.5 (4.34dB)$ (black dashed) and $s=1 (8.69dB)$ (thick black dashed). The improvements of entanglement in the TMSV gained by increasing the squeezing of the ancillary modes is small when the initial squeezing of the TMSV $r$ is small. The vertical lines from left to right indicate where the initial squeezing of the TMSV $r$ is 1dB, 3dB, 5dB, and 7dB. $\eta=\xi=1$.}
\label{prot2vss}
\end{figure}

\begin{figure}[tb]
\begin{center}
\includegraphics[width=7cm]{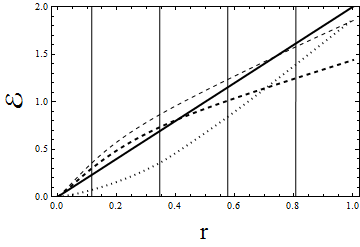}
\end{center}
\caption{Performance of protocol II (Fig.~\ref{prot2}): Logarithmic Negativity of TMSV before (solid line), and after a QND interaction for different QND interaction strengths. The QND interaction is of the form $\kappa\hat{X}\hat{P}$. The non-Gaussian ancillary modes are initially squeezed by $s=0.2 (1.74dB)$ and it is assumed that the homodyne measurement outcome is $z\approx0$. The interaction strengths are $\kappa=0.1$ (black dotted), $\kappa=0.5$ (black dashed) and $\kappa=1$ (thick black dashed). An intermediate interaction strength is preferable. The vertical lines from left to right indicate where the initial squeezing of the TMSV $r$ is 1dB, 3dB, 5dB, and 7dB. $\eta=\xi=1$.}
\label{prot2vskap}
\end{figure}

At this point the remaining covariance matrix for $A$ and $B$ and the correlations due to the $z$ measurements can be described by a $6\times6$ matrix $\mu^{(ij)}$ such that
\begin{equation}
\left(\mu^{(ij)}\right)^{-1}=\left(\begin{array}{cc}\mathcal{A}^{(ij)}&\mathcal{C}^{(ij)}\\\mathcal{C}^{T(ij)}&\mathcal{B}^{(ij)}\\\end{array}\right)
\end{equation}
where $\mathcal{A}^{(ij)}$ is a $4\times4$ matrix (modes $A$ and $B$), $\mathcal{B}^{(ij)}$ is a $2\times2$ matrix descibing the $z$ correlations, and $\mathcal{C}^{(ij)}$ contains the cross correlations. The probability of projection of modes $a$ and $b$ is given by
\begin{align}
q^{(ij)}_z=&\frac{\sqrt{\det\left(\mathcal{B}^{(ij)}-\mathcal{C}^{T(ij)}\mathcal{A}^{(ij)-1}\mathcal{C}^{(ij)}\right)}}{\pi}\nonumber\\&\times\exp\left[-\left(z,z\right)\left[\mathcal{B}^{(ij)}-\mathcal{C}^{T(ij)}\mathcal{A}^{(ij)-1}\mathcal{C}^{(ij)}\right]\left(z,z\right)^T\right]%\nonumber\\& e^{-\left(z,z\right)\left[\mathcal{B}^{(ij)}-\mathcal{C}^{T(ij)}\mathcal{A}^{(ij)-1}\mathcal{C}^{(ij)}\right]\left(\begin{array}{c}z\\z\\\end{array}\right)}
\label{qz}
\end{align}
and the Q function describing modes $A$ and $B$ reads
\begin{align}
Q^{(ij)}_{AB}\left(\alpha,\beta\right)=&\frac{\sqrt{\det\Phi^{(ij)}}}{\pi^2}e^{-\frac{-\Lambda^{(ij)}\Phi^{(ij)-1}\Lambda^{(ij)\dagger}}{4}}\nonumber\\&\times\exp\left[-\mathcal{R}^T\Phi^{(ij)}\mathcal{R}-\Lambda^{(ij)}\mathcal{R}\right]
\end{align}
where
\begin{equation}
\Phi^{(ij)}=\textbf{U}\left(\mathcal{A}^{(ij)-1}+\mathbb{1}\right)^{-1}\textbf{U}^\dagger
\end{equation}
\begin{equation}
\Lambda^{(ij)}=2\left(z,z\right)\mathcal{C}^{(ij)T}\mathcal{A}^{(ij)-1}\left(\mathcal{A}^{(ij)-1}+\mathbb{1}\right)^{-1}\textbf{U}^\dagger
\end{equation}
By defining $\Phi'^{(ij)}=\textbf{B}\left[\Phi^{(ij)}-\left(1/2\right)\mathbb{1}\right]\textbf{D}$ and $\Lambda'=\Lambda\textbf{D}$ we can write the density matrix elements as
\begin{align}
\rho_{lmnp}=&\frac{4M_{\textrm{hom}}}{\sqrt{l!m!n!p!}}\sum^1_{i,j=0}\left(-1\right)^{i+j}P^{(ij)}q^{(ij)}_z\nonumber\\&\times\sqrt{\det\Phi^{(ij)}}
e^{-\frac{-\Lambda^{(ij)}\Phi^{(ij)-1}\Lambda^{(ij)T}}{4}}H^{\{\Phi'^{(ij)},\Lambda'^{(ij)}\}}_{lmnp}
\end{align}
with
\begin{equation}
M_{\textrm{hom}}=\left(\sum^1_{i,j=0}\left(-1\right)^{i+j}P^{(ij)}q^{(ij)}_z\right)^{-1}.
\end{equation}
From this, the logarithmic negativity of modes $A$ and $B$ can be calculated as before.

In contrast to the first protocol, the second protocol is sensitive to which interactions are used. The position and momentum correlations in the TMSV are not mixed by the QND interactions and so the effects induced by interaction Hamiltonians e.g. $H^{(\kappa XP)}_{\textrm{int}}$ and $H^{(\kappa PP)}_{\textrm{int}}$ are equivalent. Correspondingly, the choice of measurement on the ancillary modes is important. If, for example, the interaction $H^{(\kappa XP)}_{\textrm{int}}$ is used, then the momentum quadratures of the ancillary modes are, by definition, unaffected but information about the TMSV is imprinted on the position quadrature distribution. If homodyne measurements on the ancillas are performed on $p$ ($\theta=\pi/2$) then the resulting outcome can reveal nothing about the state of the TMSV and the probabilities $q^{(ij)}_Z$ \eqref{qz} are independent of $\kappa$ and $r$. As nothing can be learned probabilistically about modes $A$ and $B$, procrustean entanglement concentration cannot occur. The effects on the TMSV can be transformed away by local Gaussian operations and ancillary modes and so the entanglement of the TMSV is unchanged. The above is true irrespective of whether the ancillary modes are initially squeezed in position or momentum.

If homodyne measurements were taken of the position quadratures ($\theta=0$) then  information about the TMSV will have been probabilistically imprinted on this measured distribution. If $H^{(\kappa XP)}_{\textrm{int}}$ is used then the momentum quadratures of the TMSV contain information about the momentum quadratures of the ancillary modes. That is, some noise has been added to modes $A$ and $B$ which can assist or disrupt entanglement concentration. If the ancillary modes are squeezed in momentum, then only a little noise is added to the $p$ quadratures of the TMSV. The result is an increase in entanglement dependent on $s$ and $\kappa$ (Figure \ref{prot2vss}). If the ancillary modes are squeezed in position then the momentum quadratures are anti-squeezed and so a lot of noise is added to the TMSV, having a detrimental effect on the entanglement.

As can be seen in Figures \ref{prot2vss} and \ref{prot2vskap}, successful  entanglement concentration is achieved if the ancillary modes are squeezed in $p$-quadrature, the QND interaction  is of the form $H^{(\kappa XP)}_{\textrm{int}}$, and homodyne measurement is performed in $x$-quadratures. The protocol is noticeably insensitive to ancillary mode squeezing - the logarithmic negativity of the resulting state is largely unaffected for low levels of TMSV initial squeezing.

Unlike Protocol I a weak interaction strength ruins the entanglement in the system. Similarly, if the strength is too strong then the entanglement decreases. If $\kappa\approx0.5$ then concentration successfully occurs.

\section{Discussion and Conclusions} 

We have presented here two procrustean entanglement concentrations schemes utilizing Quantum non-Demolition (QND) interactions and photon detectors. The first scheme relied upon QND interactions between Gaussian ancillary modes and a TMSV to successfully subtract photons from the TMSV as heralded by on/off detectors. We have shown how to efficiently calculate the density matrix elements of the resulting quantum state, which can then be used to calculate the logarithmic negativity of the state. This is a non-trivial task. In the asymptotic limit of the QND interaction strength $\kappa\rightarrow0$, one would find that a single photon is subtracted from each arm of the TMSV, although the probability of heralding entanglement concentration approaches zero. This is intuitively correct as the behavior mimics the protocol of \cite{Eisert2004} in which highly transmissive beamsplitters are used in place of weak QND interactions. There, as the transmittance approaches 1, one finds that a single photon is subtracted from each arm, which is the optimal outcome for entanglement concentration. The results of Browne \etal can be revisited by replacing $S_{\textrm{QND}}$ with $S_{\textrm{BS}}\left(T\right)$ in equation \eqref{gammaqnd} where $S_{\textrm{BS}}\left(T\right)$ is the symplectic operation corresponding to a beamsplitter transformation. For two modes this is
\begin{equation}
S_{\textrm{BS}}\left(T\right)=\left(\begin{array}{cccc}T&0&\sqrt{1-T^2}&0\\0&T&0&\sqrt{1-T^2}\\\sqrt{1-T^2}&0&-T&0\\0&\sqrt{1-T^2}&0&-T\\\end{array}\right).
\label{bs}
\end{equation}
We find that the method we have used for calculating density matrix elements for this protocol offers an improvement in numerical speed and efficiency over calculations in the Fock basis directly.

Double pass schemes, i.e., letting the ancillae interact with the TMSV twice, do not give any advantage in protocol I. For example $S^{(\kappa_2\hat{X}\hat{P})}_{\textrm{QND}}S^{(\kappa_1\hat{X}\hat{P})}_{\textrm{QND}}=S^{((\kappa_1+\kappa_2)\hat{X}\hat{P})}_{\textrm{QND}}$ but the logarithmic negativity is highest for weak interaction strength. By altering the interaction between passes e.g. $S^{(\kappa_2\hat{P}\hat{X})}_{\textrm{QND}}S^{(\kappa_1\hat{X}\hat{P})}_{\textrm{QND}}$ there is some advantage over a double pass scheme using the same interaction twice but this is still eclipsed by the single pass schemes.

As to detector efficiencies, at low levels of initial TMSV squeezing $r$, there can still be an increase in entanglement for weak interactions such as $\kappa=0.1$ when the detector efficiency exceeds approximately 50\%, $\eta>~0.5$. 

The second protocol relied on QND interactions to mix a TMSV with photon subtracted squeezed vacuum modes, that is with a non-Gaussian ancillae. The non-Gaussian ancillary modes are then detected after the interaction using homodyne detectors. As was stated in the previous section, the success of this scheme is dependent on the initial squeezing of the ancillary modes, the choice of interaction, and the angle $\theta$ of homodyning. It is most likely that a homodyne measurement yields a result $z=0$, and the logarithmic negativity shows an improvement so long as the measurement outcome does not stray too far from this.

The benefit of the second protocol is that the probabilistic non-Gaussian state preparation could be performed off-line and therefore the efficiency $\eta$ of the on/off detectors could be assumed to be good. Then only the homodyne detector efficiency needs to be taken into account.

Let us now consider the feasibility of the implementation of the protocols presented here in a laboratory setting. In a purely optical setup in which the TMSV is created by parametric down conversion and the QND operations are performed with beamsplitters and squeezers, protocol I is probably the easiest to implement, although other effects such as a nonzero dark count rate of the detectors would need consideration.

However, in the atomic systems of \cite{Julsgaard2001} where the entangled macroscopic spin states of two cesium gas samples represent the TMSV and off-resonant dipole interactions with strongly polarized light form the QND interactions, the vacuum ancillary modes are not true vacuum modes. They are instead  modes of heavily polarized light. The heralding of the entanglement concentration comes from detecting photons for which the interaction has altered the polarization, but this requires heavy filtering, a problem that is most likely insurmountable with current technology. Protocol II does not suffer this problem and so is the best choice in the atomic case.
The probabilities of success in both cases are comparable with those for the beamsplitter-based light schemes, already demonstrated \cite{Ourjoumtsev2007} in the laboratory. The possibility of distilling entanglement in atomic ensembles represents the main motivation for using a general QND Hamiltonian for entanglement concentration.

\section*{Acknowledgements}
We would like to thank L. Mi\v{s}ta, Jr. and J. Fiur\'a\v{s}ek for fruitful discussions. We also acknowledge the EU grant under FET-Open project COMPAS (212008) and are grateful for the support of SUPA (Scottish Universities Physics Alliance).

\appendix*

\section{Multivariate Hermite Polynomials}
\label{Hemitepolys}
In order to calculate the density matrix elements of a Gaussian state, the multivariable Hermite polynomials \cite{Bateman2} are an invaluable tool. The four dimensional Hermite polynomials are defined by
\begin{widetext}
\begin{equation}
H^{\{\Theta,\Delta\}}_{r,s,t,v}\left(y_1,y_2,y_3,y_4\right)=\left(-1\right)^{r+s+t+v}\exp\left[\textbf{y}^T\Theta\textbf{y}+\Delta\textbf{y}\right]\frac{\partial^r}{\partial y^r_1}\frac{\partial^s}{\partial y^s_2}\frac{\partial^t}{\partial y^t_3}\frac{\partial^v}{\partial y^v_4}\exp\left[-\textbf{y}^T\Theta\textbf{y}-\Delta\textbf{y}\right]
\label{Hermite}
\end{equation}
\end{widetext}
where $\textbf{y}=\left(y_1,y_2,y_3,y_4\right)^T$, $\Theta$ is a $4\times4$ matrix and $\Delta=\left(\Delta_1,\Delta_2,\Delta_3,\Delta_4\right)$. The matrix $\Theta$ is related to the inverse of the covariance matrix of the Gaussian state and the matrix $\Delta$ is related to any displacements. The density matrix elements are found by evaluating the Hermite polynomials at $y_j=0$. A recursion formula can be derived to help speed up the numerical calculation of the multivariable Hermite polynomials by directly using \eqref{Hermite} and substituting there e.g. $r+1$. This gives 
\begin{align}
H^{\{\Theta,\Delta\}}_{r+1,s,t,v}=&\Delta_1H^{\{\Theta,\Delta\}}_{r,s,t,v}\nonumber\\&-2r\Theta_{11}H^{\{\Theta,\Delta\}}_{r-1,s,t,v}\nonumber\\&-s\left(\Theta_{12}+\Theta_{21}\right)H^{\{\Theta,\Delta\}}_{r,s-1,t,v}\nonumber\\&-t\left(\Theta_{13}+\Theta_{31}\right)H^{\{\Theta,\Delta\}}_{r,s,t-1,v}\nonumber\\&-v\left(\Theta_{14}+\Theta_{41}\right)H^{\{\Theta,\Delta\}}_{r,s,t,v-1}.
\label{recursion}
\end{align}
In the above $H^{\{\Theta,\Delta\}}_{r,s,t,v}\equiv H^{\{\Theta,\Delta\}}_{r,s,t,v}\left(0,0,0,0\right)$. Similar formulae can be derived for $H^{\{\Theta,\Delta\}}_{r,s+1,t,v}$ etc. by replacing the coefficients in \eqref{recursion}. There are some symmetries which can be exploited. As the resulting density matrix is necessarily Hermitian, $H^{\{\Theta,\Delta\}}_{r,s,t,v}=H^{*\{\Theta,\Delta\}}_{t,v,r,s}$. Furthermore, if $\Delta=0$ as in protocol I or in protocol II if the homodyne projection yields $z=0$, then only Hermite polynomials for which the sum of the indices is an even number need to be calculated as whenever the sum is odd, the corresponding polynomial is zero.

%\bibliography{C:/Users/Richard/Documents/BIBTEX/References}
%\bibliographystyle{unsrt}

\end{document}